\documentclass[preprintnumbers,amsmath,amssymbm,prd]{revtex4}
\usepackage{epsfig}
\usepackage{graphicx}
\usepackage{amssymb}

\begin{document}
\title{Marginally stable Schwarzschild-black-hole-non-minimally-coupled-Proca-field bound-state configurations}
\author{Shahar Hod}
\affiliation{The Ruppin Academic Center, Emeq Hefer 40250, Israel}
\affiliation{ }
\affiliation{The Jerusalem Multidisciplinary Institute, Jerusalem 91010, Israel}
\date{\today}

\begin{abstract}
\ \ \ It has recently been revealed that, in curved black-hole spacetimes, 
non-minimally coupled massive Proca fields may be characterized by the existence of poles in their 
linearized perturbation equations and may therefore develop exponentially growing instabilities. 
Interestingly, recent numerical computations [H. W. Chiang, S. Garcia-Saenz, and A. Sang, arXiv:2504.04779] have provided compelling evidence that the onset of monopole instabilities in the composed black-hole-field system 
is controlled by the dimensionless physical parameter $\mu r_-$, where $\mu$ is the proper mass of the 
non-minimally coupled Proca field and $r_-\equiv (-2\alpha)^{1/3}r_{\text{H}}$ is the radial location of the pole [here $\alpha$ is 
the non-minimal coupling parameter of the Einstein-Proca theory and $r_{\text{H}}$ is the radius of the black-hole horizon]. 
In the present paper we use {\it analytical} techniques in order
to explore the physical properties of critical (marginally-stable) composed 
Schwarzschild-black-hole-nonminimally-coupled-monopole-Proca-field configurations. 
In particular, we derive a remarkably compact analytical formula for the 
discrete spectrum $\{\mu(r_{\text{H}},r_-;n) \}^{n=\infty}_{n=1}$ of Proca field masses which characterize the 
critical black-hole-monopole-Proca-field configurations in 
the dimensionless regime ${{r_- -r_{\text{H}}}\over{r_{\text{H}}}}\ll1$ 
of near-horizon poles. 
The physical significance of the analytically derived resonance spectrum stems from the fact that the 
critical field mass $\mu_{\text{c}}\equiv\mu(r_{\text{H}},r_-;n=1)$ 
marks the onset of instabilities in the 
Schwarzschild-black-hole-nonminimally-coupled-monopole-Proca-field system. 
In particular, composed black-hole-linearized-Proca-field configurations in the small-mass 
regime $\mu\leq\mu_{\text{c}}$ of the Proca field are stable.  
\end{abstract}
\bigskip
\maketitle

\section{Introduction}

The no-hair conjecture \cite{Whee,Car} proposed that spatially regular field configurations 
cannot be support in static equilibrium states outside the absorbing horizons of central black holes. 
Following this celebrated conjecture, a series of no-hair theorems have been established in the physics literature 
\cite{Chas,Hart,BekVec} (see also \cite{BekMay,Her1n,Hodstationary} and references therein), explicitly ruling out the existence of asymptotically flat black-hole spacetimes containing static regular configurations of scalar fields, spinor fields, and minimally coupled massive vector fields.

However, non-vacuum black-hole solutions that violate the no-hair conjecture are known to exist 
in a range of Einstein-matter field theories. In particular, it has been established that black holes 
with spatially regular horizons can support Yang-Mills hair \cite{fir,VolGal}, 
Skyrme hair \cite{hairSky}, Higgs hair \cite{hairHig1,hairHig2}, stringy hair \cite{hairstr}, 
stationary configurations of spatially regular massive bosonic fields \cite{Hod1,Hod2,Her1,Her2,Gar1}, 
scalar field configurations which are non-minimally coupled 
to the Maxwell electromagnetic invariant \cite{Hersc1,Hersc2,Hodsc1,aaa1,aaa2,Moh}, 
and electroweak hairy field configurations \cite{GV,Hodew}. 

The composed Einstein-nonminimally-coupled-Proca field theory has recently attracted much attention 
from physicists and mathematicians (see \cite{EP1,EP2,CGA} and references therein). 
In particular, it has been revealed in the physically important works \cite{EP1,EP2,CGA} 
that the perturbation equations of non-minimally coupled linearized 
Proca fields are characterized by the presence of poles whose radial locations depend on the value of the 
non-minimal coupling parameter of the theory. 

Intriguingly, it has been shown \cite{CGA} that if the pole is located outside the horizon of the 
central black hole, then the composed black-hole-massive-Proca-field system is characterized by the 
presence of a discrete family of trapped resonances that may grow exponentially in time. 
In particular, it has been demonstrated in \cite{CGA} that, for 
monopole perturbation modes \cite{Notel0} of the non-minimally coupled Proca fields, there is a pole-dependent 
critical existence-line $\mu_{\text{c}}=\mu_{\text{c}}(r_{\text{H}},r_-)$ of the Proca field mass 
that marks the boundary between stable black-hole-linearized-Proca-field systems 
and unstable black-hole-field configurations [here $r_{\text{H}}$ is the horizon radius of the 
black hole and $r_-=(-2\alpha)^{1/3}r_{\text{H}}$ is 
the radial location of the pole, where $\alpha$ is the non-minimal coupling parameter of the Einstein-Proca field 
theory, see Eq. (\ref{Eq10}) below]. 

Using direct numerical techniques, the dimensionless critical value
\begin{equation}\label{Eq1}
\mu_{\text{c}}r_-\simeq 2.8\ \ \ \ \text{for}\ \ \ \ r_-\gg r_{\text{H}}\  ,
\end{equation}
has been found in the physically interesting work \cite{CGA} 
for the composed Schwarzschild-black-hole-massive-Proca-field system in the 
regime $r_-\gg r_{\text{H}}$. 
 
As explicitly shown in \cite{CGA}, the onset of monopole instabilities in the 
Einstein-nonminimally-coupled-Proca field theory is marked by the presence of composed black-hole-massive-field `cloudy' 
configurations \cite{Notelc} which describe Schwarzschild black-hole spacetimes 
that support spatially regular configurations of linearized static Proca fields. 

The main goal of the present compact paper is to explore the physical and mathematical properties of
the composed Schwarzschild-black-hole-linearized-nonminimally-coupled-Proca-field cloudy configurations. 
Interestingly, below we shall explicitly prove that the static black-hole-monopole-field bound-state configurations 
of the non-minimally coupled Einstein-Proca field theory can be studied {\it analytically} 
in the regime
\begin{equation}\label{Eq2}
r_--r_{\text{H}}\ll r_{\text{H}}\
\end{equation} 
of near-horizon poles. 

In particular, we shall derive a remarkably compact analytical formula [see Eq. (\ref{Eq28}) below] 
for the discrete 
resonance spectrum $\{\mu(r_-;n) \}^{n=\infty}_{n=1}$ 
that characterizes the marginally-stable (static) Schwarzschild-black-hole-linearized-Proca-field 
cloudy configurations which mark the onset of monopole instabilities in the composed 
Einstein-nonminimally-coupled-Proca field theory.

\section{Description of the system}

We shall analyze the physical and mathematical properties of the critical (marginally-stable) 
Schwarzschild-black-hole-nonminimally-coupled-monopole-Proca-field 
bound-state configurations that characterize the composed Einstein-Proca field theory.

The action of the Einstein-nonminimally-coupled-Proca field theory is given by the integral expression \cite{CGA}
\begin{equation}\label{Eq3}
S=\int d^4\sqrt{-g}\Big[{{M^2_{\text{Pl}}}\over{2}}R-{{1}\over{4}}F^{\mu\nu}F_{\mu\nu}-
{{\mu^2}\over{2}}A^{\mu}A_{\mu}+{{\tilde\alpha}\over{4}}{\tilde R}^{\mu\nu\rho\sigma}F_{\mu\nu}F_{\rho\sigma}+
\beta G^{\mu\nu}A_{\mu}A_{\nu}\Big]\  ,
\end{equation}
where $F_{\mu\nu}$ is the Abelian field strength, $G_{\mu\nu}$ is the Einstein tensor, 
$R$ is the Ricci curvature scalar, and ${\tilde R}^{\mu\nu\rho\sigma}=
{1\over4}\epsilon^{\mu\nu\mu'\nu'}\epsilon^{\rho\sigma\rho'\sigma'}R_{\mu'\nu'\rho'\sigma'}$ 
is the double-dual Riemann tensor. The physical parameters $\{\mu,\tilde\alpha,\beta\}$ are 
respectively the proper mass of the Proca field and the non-trivial (non-minimal) coupling parameters of the Einstein-Proca 
theory \cite{CGA,Notealbt,Noteunits}. 

The curved line element of the supporting Schwarzschild black-hole spacetime can be expressed in the form \cite{Chan}
\begin{equation}\label{Eq4}
ds^2=-f(r)dt^2+{1\over{f(r)}}dr^2+r^2(d\theta^2+\sin^2\theta
d\phi^2)\  ,
\end{equation}
where the dimensionless metric function in (\ref{Eq4}) is given by the 
radially-dependent expression
\begin{equation}\label{Eq5}
f(r)=1-{{2M}\over{r}}\  ,
\end{equation}
where $M$ is the mass of the black hole. 
The horizon radius 
\begin{equation}\label{Eq6}
r_{\text{H}}=2M\
\end{equation}
of the central supporting black hole is determined by the root of the metric function (\ref{Eq5}).

The radial functional behavior of the non-minimally coupled monopole Proca field $\psi_{\text{M}}(r)$ 
is determined by a 
Schr\"odinger-like ordinary differential equation of the form \cite{CGA}
\begin{equation}\label{Eq7}
{{d^2\psi_{\text{M}}}\over{dy^2}}+(\omega^2-V)\psi_{\text{M}}=0\  ,
\end{equation}
where the differential relation
\begin{equation}\label{Eq8}
{{dr}\over{dy}}=f(r)\
\end{equation}
determines the tortoise coordinate $y$ of the system. 
The composed Schwarzschild-black-hole-nonminimally-coupled-monopole-Proca-field interaction potential 
in (\ref{Eq7}) is given by the (rather cumbersome) radially-dependent functional expression \cite{CGA}
\begin{equation}\label{Eq9}
V=V[r(y)]={{1-{{r_{\text{H}}}\over{r}}}\over{1-{{r^3_-}\over{r^3}}}}\cdot\Big[\mu^2-\Big(1-{{r^3_-}\over{r^3}}\Big)
\Big({{2}\over{r^2}}-{{3r_{\text{H}}}\over{r^3}}\Big)\Big]\  ,
\end{equation}
where 
\begin{equation}\label{Eq10}
r_-\equiv(-2\alpha)^{1/3}\cdot r_{\text{H}}\  
\end{equation}
is the radial location of the pole and 
\begin{equation}\label{Eq11}
\alpha\equiv {\tilde\alpha}/r^2_{\text{H}}\  
\end{equation}
is the dimensionless coupling parameter of the Einstein-Proca field theory. 

Interestingly, in the next section we shall explicitly prove that the Schr\"odinger-like ordinary differential 
equation (\ref{Eq7}) with the effective radial potential (\ref{Eq9}), which characterizes the monopole perturbation mode of the non-minimally coupled 
Einstein-Proca field theory (\ref{Eq3}), is amenable to an {\it analytical} treatment 
in the dimensionless regime $({{r_- -r_{\text{H}}})/{r_{\text{H}}}}\ll1$ of near-horizon poles. 

\section{The discrete resonance spectrum of the critical (marginally-stable) 
Schwarzschild-black-hole-nonminimally-coupled-Proca-field bound-state configurations}

As discussed in \cite{CGA}, the existence of the pole $r=r_-$ in the effective radial potential (\ref{Eq9}) implies that, 
for $r_->r_{\text{H}}$ [that is, for $\alpha<-1/2$, see Eq. (\ref{Eq10})], 
the composed black-hole-Proca-field system is characterized by two sets of 
resonant modes: an ``interior'' set which determines the linearized dynamics of the field 
in the radial region $r\in(r_{\text{H}},r_-)$ and 
an ``exterior'' set which determines the dynamics in the radial region $r\in(r_-,\infty)$. 

Static (marginally-stable) Schwarzschild-black-hole-nonminimally-coupled-monopole-Proca-field bound-state 
configurations in the interior region $r\in(r_{\text{H}},r_-)$ are characterized by the relation \cite{CGA}
\begin{equation}\label{Eq12}
\omega=0\
\end{equation}
with the physically motivated boundary condition \cite{CGA}
\begin{equation}\label{Eq13}
\psi_{\text{M}}(r=r_{\text{H}})<\infty\
\end{equation}
at the horizon of the central supporting black hole, and \cite{CGA}
\begin{equation}\label{Eq14}
\psi_{\text{M}}(r=r_-)=0\
\end{equation}
at the radial location of the pole. 

The set of equations (\ref{Eq7}), (\ref{Eq8}), (\ref{Eq9}), (\ref{Eq12}), (\ref{Eq13}), and (\ref{Eq14}) 
determines the discrete resonant spectrum $\{\mu(r_{\text{H}},r_-;n) \}^{n=\infty}_{n=1}$ 
that characterizes the critical (marginally-stable) 
black-hole-monopole-field bound-state configurations of the non-minimally coupled 
Einstein-Proca field theory (\ref{Eq3}). 
In the present section we shall study, using analytical techniques, the properties of this 
black-hole-field resonance spectrum.

In particular, we shall explicitly prove that the Schr\"odinger-like ordinary differential equation (\ref{Eq7}), 
which determines the radial functional behavior of the linearized bound-state nonminimally-coupled monopole 
Proca fields in the Schwarzschild black-hole spacetime (\ref{Eq4}), 
is amenable to an {\it analytical} treatment in the regime
\begin{equation}\label{Eq15}
x_{\text{p}}\equiv{{r_- -r_{\text{H}}}\over{r_{\text{H}}}}\ll1\
\end{equation}
of near-horizon poles. 

To this end, we first note that the radial differential equation (\ref{Eq7}) of the nonminimally-coupled 
massive Proca fields can be expressed in the form [see Eqs. (\ref{Eq7}), (\ref{Eq8}), (\ref{Eq9}), and (\ref{Eq12})]
\begin{equation}\label{Eq16}
{{d}\over{dr}}\Big[f(r){{d\psi_{\text{M}}}\over{dr}}\Big]-
{{1}\over{1-{{r^3_-}\over{r^3}}}}\cdot\Big[\mu^2-\Big(1-{{r^3_-}\over{r^3}}\Big)
\Big({{2}\over{r^2}}-{{3r_{\text{H}}}\over{r^3}}\Big)\Big]\psi_{\text{M}}=0\  .
\end{equation}

Defining the dimensionless radial coordinate 
\begin{equation}\label{Eq17}
x\equiv{{r -r_{\text{H}}}\over{r_{\text{H}}}}\  ,
\end{equation}
where $x\in[0,x_{\text{p}}]$ in the interval $r\in[r_{\text{H}},r_-]$, 
one finds the near-horizon relations
\begin{equation}\label{Eq18}
{{d}\over{dr}}\Big[f(r){{d\psi_{\text{M}}}\over{dr}}\Big]=
{{1}\over{r^2_{\text{H}}}}\cdot\Big[x{{d^2\psi_{\text{M}}}\over{dx^2}}+{{d\psi_{\text{M}}}\over{dx}}\Big]\cdot[1+O(x)]\  ,
\end{equation} 
\begin{equation}\label{Eq19}
{{\mu^2}\over{1-{{r^3_-}\over{r^3}}}}={{\mu^2}\over{3(x-x_{\text{p}})}}\cdot[1+O(x)]\  ,
\end{equation}
and
\begin{equation}\label{Eq20}
{{2}\over{r^2}}-{{3r_{\text{H}}}\over{r^3}}=-{{1}\over{r^2_{\text{H}}}}\cdot[1+O(x)]\
\end{equation}
in the radial region $x\leq x_{\text{p}}\ll1$ [see Eqs. (\ref{Eq15}) and (\ref{Eq17})]. 

Taking cognizance of Eqs. (\ref{Eq18}), (\ref{Eq19}), and (\ref{Eq20}) and defining 
the dimensionless mass-radius parameter
\begin{equation}\label{Eq21}
{\bar\mu}\equiv \mu r_-\
\end{equation}
of the composed black-hole-field system, one finds that, 
in the near-horizon regime (\ref{Eq15}) with \cite{Noteblw} 
\begin{equation}\label{Eq22}
{{{\bar\mu}^2}\over{x_{\text{p}}}}\gg1\  ,
\end{equation}
the differential equation (\ref{Eq16}) for the static (marginally-stable) monopole Proca configurations 
can be written in the form 
\begin{equation}\label{Eq23}
x{{d^2\psi_{\text{M}}}\over{dx^2}}+{{d\psi_{\text{M}}}\over{dx}}+{{{\bar\mu}^2}\over{3(x_{\text{p}}-x)}}\psi_{\text{M}}=0\  .
\end{equation} 

Interestingly, and most importantly for our analysis, the radial differential equation (\ref{Eq23}) 
can be solved {\it analytically}. 
In particular, the mathematical solution of the monopole equation (\ref{Eq23}) 
that respects the physically motivated 
boundary condition (\ref{Eq13}) of regular Proca field configurations at the horizon of the 
central supporting Schwarzschild black hole is given by the radial functional expression
\cite{Abram,Morse,Notenm}
\begin{equation}\label{Eq24}
\psi_{\text{M}}(x)={_2F_1}\Big(-{{{\bar\mu}}\over{\sqrt{3}}},{{{\bar\mu}}\over{\sqrt{3}}};1;{{x}\over{x_{\text{p}}}}\Big)\ ,
\end{equation}
where ${_2F_1}(a,b;c;z)$ is the hypergeometric function \cite{Abram,Morse}. 

Taking cognizance of the boundary condition (\ref{Eq14}) at the radial location $x=x_{\text{p}}$ 
of the pole and using the functional relation \cite{Abram,Morse}
\begin{equation}\label{Eq25}
{_2F_1}(-a,a;1;1)={{\sin(\pi a)}\over{\pi a}}\
\end{equation}
for the hypergeometric function, one obtains the resonance condition
\begin{equation}\label{Eq26}
\sin\Big({{\pi{\bar\mu}}\over{\sqrt{3}}}\Big)=0\
\end{equation}
that characterizes the composed black-hole-monopole-field system. 
The analytically derived equation (\ref{Eq26}) yields the remarkably compact 
discrete resonance spectrum \cite{Notenn0,Notest}
\begin{equation}\label{Eq27}
{\bar\mu}=\sqrt{3}\cdot n\ \ \ \ ; \ \ \ \ n=1,2,3,...
\end{equation}
for the critical (marginally-stable) composed Schwarzschild-black-hole-nonminimally-coupled-Proca-field bound-state configurations. 
 
\section{Summary}

It has recently been revealed in the physically interesting works \cite{EP1,EP2,CGA} that, 
in the composed Einstein-Proca field theory (\ref{Eq3}), 
the effective interaction potential that appears in the linearized perturbation equations 
of the non-minimally coupled massive fields is characterized by the presence of a pole 
whose radial location $r_-=r_-(\alpha)$ depends on the value of the non-minimal coupling 
parameter $\alpha$ of the theory [see Eq. (\ref{Eq10})]. 

In particular, it has been shown \cite{CGA} that if the pole is located outside the horizon of the 
central black hole, $r_->r_{\text{H}}$, then the composed black-hole-field system has 
a discrete family of trapped resonant modes which are supported in the radial interval $[r_{\text{H}},r_-]$ and 
may grow exponentially in time. 

Interestingly, based on direct numerical computations, it has been nicely demonstrated in \cite{CGA}
that the onset of monopole Proca instabilities in the Einstein-Proca field theory (\ref{Eq3}) is marked by the presence of cloudy 
black-hole-field bound-state configurations which are composed of a central black hole that supports marginally-stable (static) 
linearized massive Proca fields. 

In the present paper we have studied, using {\it analytical} techniques, the physical and mathematical properties of the
composed Schwarzschild-black-hole-nonminimally-coupled-monopole-Proca-field cloudy 
configurations in the regime $r_--r_{\text{H}}\ll r_{\text{H}}$ of near-horizon poles 
[see Eq. (\ref{Eq15})]. 
In particular, we have derived the remarkably compact formula [see Eqs. (\ref{Eq21}) and (\ref{Eq27})] 
\begin{equation}\label{Eq28}
\mu r_-=\sqrt{3}\cdot n\ \ \ \ ; \ \ \ \ n=1,2,3,...
\end{equation}
for the discrete resonance spectrum of the 
dimensionless mass-radius parameter which characterizes the marginally-stable (static) 
black-hole-field cloudy configurations. 

Finally, it is worth emphasizing the fact that the physical significance of the analytically derived 
resonance spectrum (\ref{Eq28}) stems from the fact that, in the dimensionless regime (\ref{Eq15}), the 
critical field mass 
\begin{equation}\label{Eq29}
\mu_{\text{c}}\equiv\mu(r_-;n=1)={{\sqrt{3}}\over{r_-}}\
\end{equation}
marks the onset of instabilities in the composed Schwarzschild-black-hole-monopole-Proca-field 
configurations. 
In particular, composed Schwarzschild-black-hole-linearized-Proca-field systems 
are stable in the small-mass regime $\mu\leq\mu_{\text{c}}$ of the 
Proca field. 

\bigskip
\noindent
{\bf ACKNOWLEDGMENTS}
\bigskip

This research is supported by the Carmel Science Foundation. I would
like to thank Yael Oren, Arbel M. Ongo, Ayelet B. Lata, and Alona B.
Tea for helpful discussions.



\begin{thebibliography}{99}
\bibitem{Notel0} Note that the monopole Proca mode is characterized by the angular harmonic index $l=0$.
\bibitem{Notelc} The term `linearized clouds' is commonly used in the physics literature (see \cite{Hod1,Hod2,Her1,Her2,Gar1}) to describe matter configurations of static or stationary 
fields which are linearly coupled to central supporting black holes.
\bibitem{Notealbt} Note that the non-minimal coupling parameters $\{\tilde\alpha,\beta\}$ of the Einstein-Proca 
theory (\ref{Eq3}) have mass-dimension $-2$ and $0$ respectively \cite{CGA}.
\bibitem{Noteunits} Here $M_{\text{Pl}}$ is the Planck mass. 
We shall henceforth use natural units in which $G=c=1$.
\bibitem{Notenm} Note that the normalization pre-factor of the linearized non-minimally coupled monopole 
Proca field (\ref{Eq24}) is chosen, without loss of generality, to be $1$.
\bibitem{Notenn0} Note that the assumption (\ref{Eq22}) implies $n>0$ in the discrete resonance spectrum (\ref{Eq27}).
\bibitem{Notest} Note that the Proca field masses (\ref{Eq27}), which characterize 
the critical (marginally-stable) black-hole-monopole-field configurations in the dimensionless regime (\ref{Eq15}), 
respect the assumed strong inequality (\ref{Eq22}).
\end{thebibliography}
\end{document}